# Modified uni-travelling-carrier photodiodes with 206 GHz bandwidth and 0.81 A/W external responsivity

Linze Li[1,2†], Tianyu Long[1,2†], Xiongwei Yang[3,5,6†], Zhouze Zhang[1],
Luyu Wang[1,2], Jingyi Wang[1,2], Mingxu Wang[3,5,6], Juanjuan Lu[1],
Jianjun Yu[3,4,5,6*], Baile Chen[1*]

[1]School of Information Science and Technology, ShanghaiTech University, Shanghai, 201210, China.
[2]Fastphide Photonics, Shanghai, 201210, China.
[3]China State Key Laboratory of ASIC and System, Fudan University, Shanghai, 200433, China.
[4]Purple Mountain Laboratories, Nanjing, 211111, China.
[5]School of Information Science and Technology, Fudan University, Shanghai, 200433, China.
[6]Key Laboratory for Information Science of Electromagnetic Waves, Fudan University, Shanghai, 200433, China.

*Corresponding author(s). E-mail(s): jianjun@fudan.edu.cn ; chenbl@shanghaitech.edu.cn ;
Contributing authors: lilz@shanghaitech.edu.cn ; longty2022@shanghaitech.edu.cn; xwyang22@m.fudan.edu.cn ; zhangzhz2023@shanghaitech.edu.cn ; wangly4@shanghaitech.edu.cn ; wangjy8@shanghaitech.edu.cn ; 21110720068@m.fudan.edu.cn ; lujj2@shanghaitech.edu.cn ;
†These authors contributed equally to this work.

**Abstract**

The accelerating demand for wireless communication necessitates wideband, energy-efficient photonic sub-terahertz (sub-THz) sources to enable ultra-fast data transfer. However, as critical components for THz photonic mixing, photodiodes (PDs) face a fundamental trade-off between bandwidth and quantum efficiency, presenting a major obstacle to achieving high-speed performance with high optoelectronic conversion efficiency. Here, we overcome this challenge by demonstrating an InP-based, waveguide-integrated modified uni-traveling carrier photodiode (MUTC-PD) with bandwidth exceeding 200 GHz and a bandwidth-efficiency product (BEP) surpassing 130 GHz. Incorporating a spot-size converter (SSC) together with optimized electric-field distribution, balanced carrier transport, and minimized parasitic capacitance, the device achieves a 3-dB bandwidth of 206 GHz and an external responsivity of 0.81 A/W, setting a new BEP benchmark. Packaged with WR-5.1 waveguide output, it delivers radio-frequency (RF) power exceeding -5 dBm across the 127–185 GHz frequency range. As a proof of concept, we achieved a wireless transmission of 54 meters with a single-line rate of up to 120 Gbps, leveraging photonics-aided technology without requiring a low-noise amplifier (LNA). This work establishes a pathway to significantly enhance optical power budgets and reduce energy consumption, presenting a transformative step toward high-bandwidth, high-efficiency sub-THz communication systems and next-generation wireless networks.

**Keywords:** modified uni-traveling-carrier(MUTC), THz wireless communications, high-speed photodetectors, THz generation

# 1 Introduction

Driven by emerging real-time services like telemedicine, autonomous driving, and virtual reality [1, 2], the surge in wireless communication traffic has saturated traditional microwave frequency bands, prompting a shift towards the sub-terahertz (sub-THz) spectrum [3, 4]. Photonic-assisted THz technologies have emerged as a compelling alternative to all-electronic methods, offering unparalleled advantages such as broad frequency bandwidth, tunability, and intrinsic stability [5]. These attributes make photonic approaches particularly suited to addressing the high-capacity demands of future communication systems. At the heart of photonic THz-wave generation are ultra-fast photodiodes (PDs), which serve as the critical interface for optical-to-electrical conversion through photo-mixing. The performance of these devices dictates the system's maximum operating frequency, dynamic range, and overall efficiency, making their development pivotal to advancing THz communication technologies [6, 7].

However, existing high-speed PDs face limitations in photoelectric conversion efficiency near the sub-THz range, creating a dependency on high input optical power and post-amplification before transmission [8, 9]. This dependence can increase energy consumption and risk nonlinear distortions in practical use. To overcome these challenges, PDs must achieve high saturation power, high responsivity, and wide bandwidth concurrently to enhance the performance and efficiency of sub-THz systems.

Among various types of ultra-fast photodiode, the uni-traveling-carrier photodiode (UTC-PD) has exhibited the highest output power in sub-THz regime, with continuous structural modifications since its introduction in 1997 [10]. These pioneering works effectively address the carrier screening issue [11] by utilizing the fast electrons from the InP platform as active carriers in the collector [12, 13]. Surface-illuminated modified UTC-PDs (MUTC-PDs) have demonstrated the ability to achieve THz-level bandwidths of 100-330 GHz (refs. [14–18]) and high output power by thinning the absorber, reducing PD sizes and incorporating inductance peaking. However, this miniaturization leads to reduced optical absorption and considerable coupling losses, consequently decreasing efficiency and responsivities.

In contrast, waveguide photodiodes (WG-PDs) offer a promising solution to the inherent trade-off between efficiency and speed by laterally coupling light into the active region through an optical waveguide [19–21]. Recent advancements report a record 3-dB bandwidth exceeding 220 GHz and a responsivity of 0.24 A/W for InP-based evanescently coupled waveguide MUTCPDs [21]. And high RF power close to 0 dBm has also been demonstrated at 200 GHz [22] and 240 GHz [23] using waveguide MUTC-PDs, achieved through the traveling-wave structure and electric field shaping. For sub-THz speed, comprehensive engineering of carrier transport and RC response is needed. However, reducing the PD size to below 30 μm² [20, 21]—typically preferred for better RC response—introduces challenges in achieving efficient coupling due to the corresponding reduction in the absorption length. This trade-off between efficiency and RC time constant has led to previously reported ultra-fast WG-PDs facing a persistent 55 GHz (110 GHz × 50%) bottleneck [24] in bandwidth-efficiency product (BEP). Therefore, developing a UTC-PD with an enhanced BEP, higher speed, and the ability to serve as a photonic sub-THz source with low power consumption, compact design, and wide bandwidth remains a critical and sought-after goal.

In this work, we demonstrate waveguide-integrated MUTC-PDs with a record-breaking BEP exceeding 130 GHz, effectively doubling the highest BEP reported to date [24]. This outstanding performance is achieved through several synergistic strategies: fine-tuning the electric field distribution, balancing carrier transport, and incorporating benzocyclobutene (BCB) beneath the electrodes to minimize parasitic capacitance, while simultaneously integrating a spot-size converter (SSC) to enhance external responsivity and overall device efficiency. Our design achieves an unprecedented 3-dB bandwidth of 278 GHz for 2 × 7 μm² PDs, the highest reported among all III-V WG-PDs. Meanwhile, the 2 × 15 μm² devices set a new BEP record of 133.5 GHz (206 GHz × 64.8%), with bandwidths exceeding 200 GHz and an external responsivity of 0.81 A/W. To showcase the device's capability in data transmission systems, we developed a WR-5.1 waveguide-packaged MUTC-PD module, achieving maximum output power exceeding -5 dBm between 127 GHz and 185 GHz, with power maintained above -10 dBm across the entire G-band. Furthermore, leveraging photonics-aided technology and eliminating the need for a low-noise amplifier (LNA), we successfully achieved a 54-meter wireless link at data rates up to 120 Gbps, highlighting the device's exceptional capabilities for next-generation, high-capacity wireless networks.

# 2 Results

## 2.1 Device structure and design

Fig.1a shows the concept and structure of the proposed waveguide-integrated PD chip, which includes an SSC, a MUTC-PD structure and a coplanar waveguide (CPW). Fig.1b shows the corresponding scanning electron microscope (SEM) image. The 2-step SSC is utilized to achieve efficient optical coupling within compact coupling length, ensuring both high responsivity and fast RC response simultaneously. The first section is a 1.5-μm-thick diluted waveguide, designed to facilitate fiber-to-waveguide coupling through mode field matching [25, 26]. For an input Gaussian beam with a 2-μm diameter spot size, this structure achieves a simulated coupling loss as low as 0.2 dB. The second section consists of two InGaAsP optical matching layers, which provide a gradual increase of the optical refractive index from the diluted waveguide to the PD's active region, thereby enhancing upward light guidance into the active area as small as 30 μm². The external responsivity of the 2 × 15 μm² device was simulated with respect to different matching layer thickness, with $SiO_2$ on the waveguide facet serving as the antireflection coating, as shown in Fig.1f. The responsivity of 0.823 A/W can be obtained when the thickness of the first matching layer (Q1.03) and the second matching (Q1.33) layer is 210 nm and 320 nm respectively. Fig.1g shows the simulated optical intensity as light propagates through the device.

The cross section of the MUTC structure is shown in Fig.1c, which contains a P-doped partially depleted InGaAs absorption layer (165 nm) positioned between an InP diffusion block layer and an InP drift layer. A 30 nm cliff layer with doping level of $2 \times 10^{17}$ cm$^{-3}$ is inserted in the drift layer to enable rapid electron transport by shaping the electric field. Fig.1d shows the simulated electric field distribution for the 2 × 15 μm² device at various photocurrent levels. The heavily doped cliff layer effectively boosts the electric field across the InGaAs/InP interfaces (blue region), aiding electron movement across the heterogeneous inter-facial barrier and thereby alleviating the current blocking effect [27]. It also maintains an electric field of 20-50 kV/cm throughout the entire e-drift layer (grey region) at photocurrent up to 10 mA, facilitating electron velocity overshoot [28] and ensuring a short electron transit time. Additionally, this electric field regulation helps counteract space charge effects at high currents, leading to improved high-power performance. To further reduce the carrier transit time, the depleted absorber is designed to be 100 nm thick, which is noticeably thicker than the 30 nm thickness typically used in the reported MUTC-PDs with ultra-wide bandwidth [17, 18, 29]. This design effectively reduces the proportion of diffusion current from the p-type absorber, thereby allowing more electrons to be transported efficiently with the high electric field in the depletion region. Fig.1e illustrates the calculated transit-limited bandwidth as the thickness of the depleted absorber ($W_d$) varies from 50 to 150 nm. Here, the energy-balance model is employed to characterize non-stationary electron transport dynamics [30], ensuring accurate analysis of the MUTC-PD's performance. The bandwidth initially increases significantly due to the reduced electron diffusion time. When $W_d$ exceeds 100 nm, the increased contribution of slower-responding hole currents in the depleted absorption region starts to slow down the overall transit process. Consequently, a thickness of 100 nm is chosen to balance the response of electrons and hole currents, leading to an transit-limited bandwidth of 416 GHz. Moreover, thicker $W_d$ also helps small-size devices maintain acceptable junction capacitance.

Apart from transit time and junction capacitance, parasitic parameters are another critical factor affecting bandwidth. To address this, the introduction of a low dielectric constant material, such as BCB, beneath the CPW structure can effectively reduce the electrode parasitic capacitance ($C_{ep}$) [21], thereby enhancing the RC-time-limited bandwidth. As shown in Fig.1h (table format) and Fig.1i, the 3-dB bandwidth of the 2 × 15 μm² device, accounting for both RC and transit frequency responses, along with the $C_{ep}$, was simulated for various BCB thicknesses. The implementation of 1 μm thick BCB significantly enhanced the 3-dB bandwidth by approximately 30 GHz compared with the traditional scheme, where CPW electrodes are deposited directly on the InP substrate. As the BCB thickness increased to 5 μm, the device achieved a 3-dB bandwidth of 205 GHz, with only marginal improvements observed with further increases in thickness. Notably, as $C_{ep}$ decreases, a slight peak appears in the frequency response due to the resonance effect [18] introduced by the intrinsic inductance of the CPW electrode, which also contributes to extending the bandwidth. Considering both the bandwidth enhancement and fabrication feasibility, the BCB thickness was ultimately controlled within the range of 5 to 8 μm. The complete simulation process, utilizing "field-circuit co-simulation", is provided in Supplementary Note 2.

## 2.2 Characterization of PDs

The dark current characteristics for devices with varying lengths are depicted in Fig.2a. A typical dark current is observed to be around 200 pA at a -2 V bias. In Fig.2b, the measured photocurrents as a function of incident optical power are presented. At a wavelength of 1550 nm, devices with lengths of 7 μm, 10 μm, and 15 μm exhibit external responsivities of 0.29 A/W, 0.53 A/W, and 0.81 A/W, respectively, determined from the slope of the linear fit. Notably, no saturation is observed up to an incident optical power of 10 mW. Fig.2c shows the responsivities of the 2 × 15 μm² device with respect to the vertical position of the fiber under different polarization states. The highest responsivity is obtained under TE polarization, with a polarization-dependent loss (PDL) of 1.3 dB. By shifting the optical fiber downward by 0.5 μm, an extremely low PDL of -0.04 dB is achieved while maintaining a high responsivity of 0.7 A/W, ensuring minimal signal distortion and stable performance in polarization-sensitive applications. The simulated external responsivities and corresponding PDL are illustrated in Supplementary Note 3, showing strong agreement with the experimental results.

In Fig.2d-f, the frequency response of devices with lengths of 7 μm, 10 μm, and 15 μm is depicted at a -1 V bias voltage, corresponding to photocurrents of 1.5 mA, 2.2 mA, and 3 mA, respectively. The 2 × 7 μm² device exhibits an ultra-wide bandwidth of 278 GHz, showcasing the highest performance in terms of speed among waveguide-type PDs. The 2 × 10 μm² and 2 × 15 μm² devices, with larger active areas, show bandwidth of 243 GHz and 206 GHz, respectively. The reduction in bandwidth is attributed to the growing junction capacitance, indicating that the RC time constant is the major limiting factor. Nonetheless, thanks to effective parasitic capacitance optimization, the largest device still maintains a bandwidth exceeding 200 GHz and achieves a record-high BEP of 133.5 GHz (206 GHz × 64.8%).

At the center frequency of the G-band (180 GHz), the saturation performance of PDs with various lengths is assessed by measuring the RF output power as a function of the average photocurrent, as shown in Fig.2g. Under -1.5 V bias, devices with lengths of 7 μm, 10 μm and 15 μm reach output powers of -8.2 dBm, -4.1 dBm and 0 dBm, respectively. Shorter PDs achieve higher output power before saturation due to their superior RC response. However, the 2 × 15 μm² device sustains a much higher saturation photocurrent of 9.3 mA, ultimately delivering greater overall output power, which is comparable to that of ultrafast UTC-PDs in the G-band [16, 31, 32]. This is attributed to the more homogeneous optical intensity distribution in longer devices. Fig.2h illustrates the output power of the 2 × 15 μm² device under different bias voltages at 220 GHz. Increasing the bias voltage mitigates carrier accumulation by enhancing the electric field, thereby increasing the saturation current. The device exhibits the best saturation performance at -1.5 V bias, achieving a peak output power of -2.6 dBm. However, at reverse biases beyond -1.5 V, the excessively high electric field in the drift layer significantly reduces the relaxation time for velocity overshoot [33], reducing the bandwidth and consequently lowering the output power. With its record BEP and substantial output power up to 220 GHz, the 2 × 15 μm² device is particularly well-suited for G-band THz communication applications. For frequencies beyond the G-band, the saturation performance of PDs with various lengths is detailed in Supplementary Note 6, with the highest output powers of -5.4 dBm at 250 GHz and -7 dBm at 280 GHz, respectively, below the state of the art [23, 34]. The considerable power drop stems from the slight peaking effect, which flattens the frequency response within the G-band at the cost of a more rapid decline in higher frequency bands. To further improve output power beyond the G-band, strategies such as lowering CPW inductance [17] and enhancing the electric field across junction region [23] can be employed.

## 2.3 Packaged module and THz wireless communication

As the operating frequency exceeds 100 GHz, PDs packaged with coaxial connectors face challenges in achieving low-loss output, primarily due to dielectric losses, higher-order modes, and electromagnetic interference. In contrast, waveguides provide low-loss output over a wide frequency range [35–38], while also enabling seamless integration with other standard waveguide interface devices, such as antennas, amplifiers, and mixers.

To meet the demand for high-efficiency sub-THz wireless communication systems, we have designed a WR-5.1 waveguide packaged module, as illustrated in Fig.3a. In this design, light is coupled into the SSC using a cost-effective solution that includes a single-mode fiber, collimator, and microlens. The PD electrodes are connected to a 50 μm thick quartz passive circuit via wire bonding, which integrates a transition from the CPW to a modified radius stub probe to facilitate low-loss transmission of RF signals. It also enables the excitation of the $TE_{10}$ mode in the rectangular waveguide. A step-impedance low-pass filter is incorporated to prevent RF signal leakage while providing a DC bias to the PD. Through careful design and optimization of the quartz passive circuit, we achieve an insertion loss of approximately 0.5 dB across the frequency range of 140–220 GHz, ensuring minimal RF loss. A back-to-back module is employed to characterize the loss properties, with detailed descriptions provided in Supplementary Note 7.

After packaging the 2 × 15 μm² device without an anti-reflective (AR) coating, the responsivity of the module is measured to be 0.5 A/W. The frequency response of the packaged module under different photocurrents is shown in Fig.3b. At a photocurrent of 8.5 mA, the output power exceeds -5 dBm from 127 GHz to 185 GHz, and remains above -10 dBm across the entire G-band. These results demonstrate the substantial potential of the MUTC-PD module for applications in short-range radar, indoor communication, long-range transmission, and other related scenarios.

Fig.3c-e present experimental results of photonics-aided sub-THz wireless transmission over 54 m using the waveguide-packaged MUTC-PD module. The experimental setup is described in detail in Supplementary Note 8. Fig.3c shows the error vector magnitude (EVM) curve for a 20 Gbaud quadrature phase shift keying (QPSK) signal transmitted over 54 meters in the 135 GHz to 165 GHz frequency range. The EVM remains below 15%, well below the 17.5% threshold for QPSK signals, indicating that the module provides sufficient saturated output power for 54-meter wireless transmission without the need for THz amplification. Fig.3d shows the relationship between EVM and baud rate for the QPSK signal. At a photocurrent of 4 mA, the module supports the generation and transmission of a 30 Gbaud QPSK signal over 54 meters. Although the module can support a wider bandwidth, the mixer used in the experimental setup limits the achievable transmission rate, preventing higher rates from being demonstrated. The 25 Gbaud QPSK constellation diagram in Fig.3d shows near-error-free transmission. Further transmission experiments are conducted with 30 Gbaud QPSK and 16quadrature amplitude modulation (QAM)

signals, centered at 150 GHz. Fig.3e shows the bit error rate (BER) as a function of photocurrent after 54 meters of wireless transmission. As photocurrent increases, the RF signal power also increases, resulting in a reduction in BER. At a photocurrent of 5 mA, the BERs for 30 Gbaud QPSK and 16-QAM signals are $3\times10^{-5}$ and $2.3\times10^{-2}$, respectively—both well below the hard-decision forward error correction (FEC) ($3.8\times10^{-5}$) and soft-decision FEC ($2.4\times10^{-2}$) thresholds. These results demonstrate that the transmission system can support a line rate of 120 Gbps over 54 m without THz amplifiers, highlighting the substantial improvements in output power and efficiency achieved with the designed MUTC-PD module.

# 3 Discussion

Fig.4 provides an overview of state-of-the-art waveguide-coupled and surface illuminated PDs, with detailed parameters summarized in Supplementary Note 9. While surface-illuminated PDs have demonstrated impressive speed performance, reaching up to 330 GHz, their responsivity significantly decreases as the bandwidth increases, particularly for PDs with bandwidths exceeding 200 GHz. WG-PDs, on the other hand, decouple the optical coupling direction from the carrier transport path, enabling BEP between 37 and 55 GHz by realizing both efficient absorption and short carrier transit times. In this work, the 2 × 15 μm² device demonstrates an enhanced bandwidth of 206 GHz and an external responsivity of 0.81 A/W, resulting in a breakthrough BEP of 133.5 GHz (206 GHz × 64.8%). Additionally, the 2 × 7 μm² device achieves a record bandwidth of 278 GHz among all waveguide-coupled PDs. This exceptional performance can be attributed to the integration of a SSC, optimization of carrier transport, and the reduction of parasitic capacitance.

Table 1 summarizes the latest progress in short-distance photonics-aided transmission systems within the 110 GHz to 170 GHz frequency range, detailing parameters such as center frequency, PD type, LNA gain, transmission rate, and distance. Here, an LNA gain of 0 dB indicates that no THz amplifier was used in the transmission system. As shown in Table 1, previously reported systems without an LNA can only achieve bit rates of up to 60 Gbps with a short transmission distance of 3 meters, limiting their practical applications. While the inclusion of an LNA improves both the transmission rate and distance, it comes at the cost of higher energy consumption and risk of nonlinear distortions. In contrast, we experimentally demonstrate a single-line rate of 120 Gbps over a 54-meter transmission distance without the use of THz amplifiers. At a bias voltage of -1.5 V and a photocurrent of 5 mA, the corresponding RF power and efficiency of our MUTC-PD module are sufficient to support wireless transmission over 54 meters, achieving an improved data rate.

In conclusion, we have demonstrated a waveguide-integrated MUTC-PD with an ultrahigh bandwidth exceeding 200 GHz and a record BEP of 133.5 GHz (206 GHz × 64.8%). After packaged with a WR-5.1 waveguide output, the device can support 54-meter wireless transmission at a single line rate of up to 120 Gbps, utilizing photonics-aided technology without any LNA. This work emphasizes the potential of InP photonics integrated circuits for efficient THz generation, paving the way for next-generation wireless communication systems with enhanced optical power budgets.

# Acknowledgements

B. Chen acknowledges support from the National Key Research and Development Program of China (grant no. 2018YFB2201000) and the National Natural Science Foundation of China (grant no. 61975121). J. Yu acknowledges support from the National Natural Science Foundation of China (grant nos. 62127802, 62331004, 62305067, U24B20142, U24B20168, and 62427815). The PD chips were fabricated with support from the ShanghaiTech Material and Device Lab (SMDL).

## Author contributions

B. Chen conceived the idea. L. Li, Z. Zhang, and B. Chen designed the device epi-layer structure. L. Li, T. Long, L. Wang, Z. Zhang, and J. Wang jointly fabricated the devices and participated in the testing process. T. Long designed and packaged the WR-5.1 module and tested its performance. X. Yang, T. Long, Z. Zhang and M. Wang conducted the terahertz communication experiment supervised by J. Yu. L. Li, T. Long, X. Yang, Z. Zhang, L. Wang, J. Wang, M. Wang, J. Lu, and B. Chen primarily wrote the manuscript. B. Chen initiated the collaboration and supervised the project.

## Competing interests

The authors declare no competing interests.

**Table 1 Comparison of short-distance photonics-aided transmission systems**

| Year/reference | PD type | Frequency (GHz) | LNA gain (dB) | Bit Rate (Gbps) | Distance (m) |
|---|---|---|---|---|---|
| 2022[47] | UTC-PD | 135 | 0 | 60 | 3 |
| 2022[48] | UTC-PD | 130 | 0 | 60 | BTB |
| 2022[49] | UTC-PD | 120 | 30 | 55 | 200 |
| 2023[50] | UTC-PD | 112 | 30 | 57.2 | 13 |
| 2024[51] | UTC-PD | 120 | 30 | 72 | 70 |
| 2024[52] | MUTC-PD | 150 | 18 | 75 | 1 |
| 2024[53] | MUTC-PD | 131 | 18 | 120 | 1 |
| This work | MUTC-PD | 150 | 0 | 120 | 54 |

LNA, low noise amplifier; BTB, back to back.

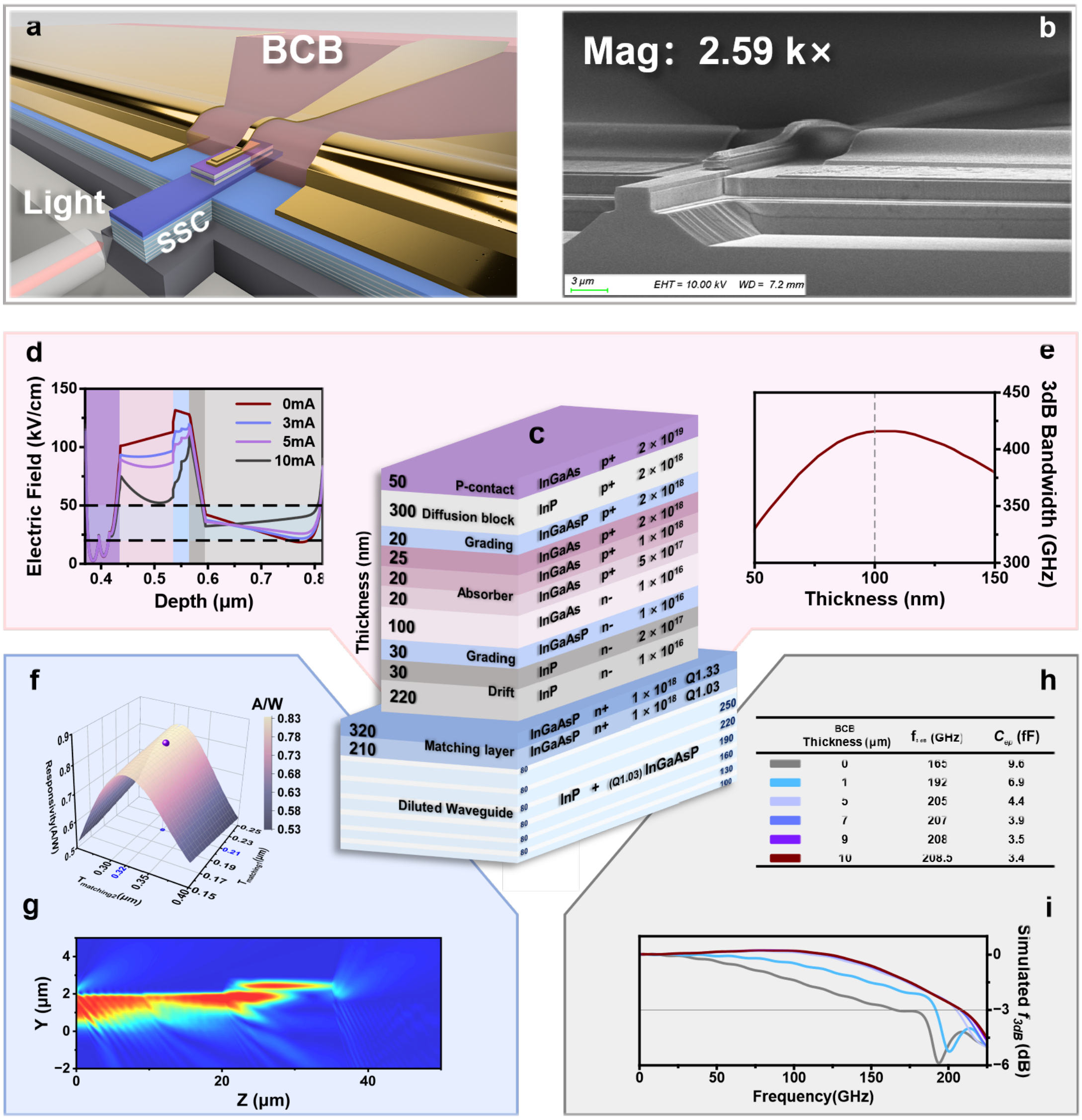

**Fig. 1 The terahertz waveguide-integrated MUTC-PD**. **a**, Schematic of the device. The integrated SSC efficiently couples the incident optical signal into the active area of MUTC-PD. The RF output signal is transmitted through a CPW taper to the contact pads of the PD. **b**, SEM image of the fabricated device. **c**, Epi-layer structure of the SSC and the MUTC-PD structure. The epitaxial growth begins with a diluted waveguide composed of six InGaAsP layers, with thicknesses increasing from 100 nm at the bottom to 250 nm at the top in 30-nm increments, interspersed with 80-nm-thick InP layers. Two InGaAsP layers above the diluted waveguide serve as the optical matching layers. For the MUTC-PD structure, a 165 nm P-doped InGaAs absorption layer is partially depleted, with a 65 nm undepleted absorption region featuring a step-grading profile ($5 \times 10^{17}$ cm$^{-3}$,$1 \times 10^{18}$ cm$^{-3}$, and $2 \times 10^{18}$ cm$^{-3}$) to induce a quasi-electric field that assists electron transport. The top 300 nm p-doped InP layer acts as the diffusion block layer, while the lower 250 nm slightly n-doped InP layer serves as the drift layer. InGaAsP quaternary layers are designed at InGaAs/InP hetero-junction interfaces to smooth the energy band discontinuities. **d**, Simulated electric field distributions at various photocurrent for the 2 × 15 μm² device under -1.5 V bias. **e**, Simulated transit-limited bandwidth as a function of $W_d$. **f**, Simulated external responsivity versus thickness of the first (Q1.03) and the second (Q1.33) matching layers for the 2 × 15 μm² device. **g**, Simulated optical intensity as light propagates through the structure. **h**, The detailed data of Simulated 3-dB bandwidth and the $C_{ep}$ of the 2 × 15 μm² device with varying thickness of BCB. **i**, Simulated frequency response of the 2 × 15 μm² device with varying thickness of BCB.

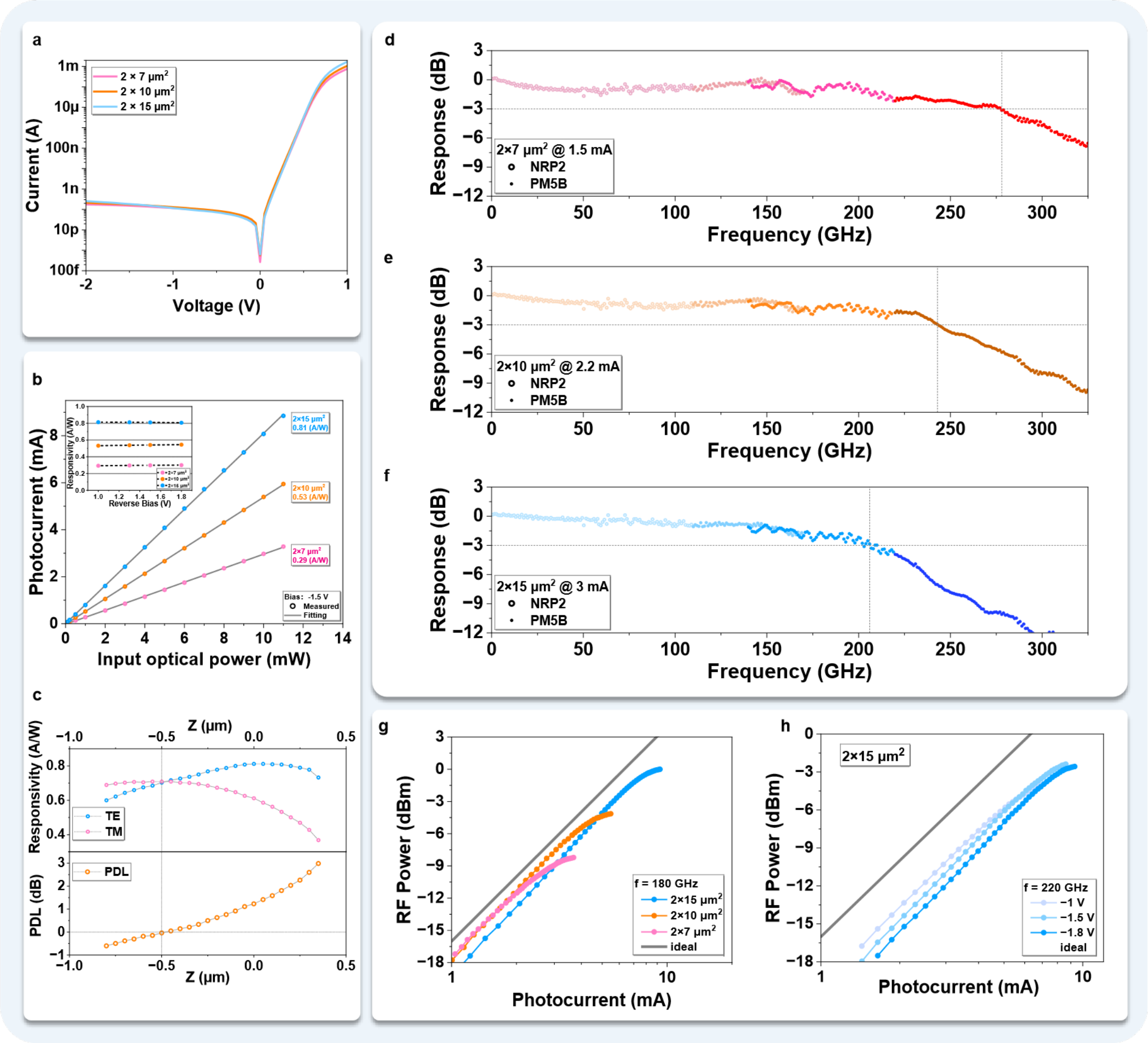

**Fig. 2 Measured static and dynamic performance of PDs**. **a**, Dark current versus bias voltage characteristic for the devices with different lengths. **b**, Measured photocurrents as a function of incident optical power for devices with different lengths under -1.5 V bias. The corresponding external responsivities are determined from linear fit. The inset shows the fitted external responsivities under different bias voltages. **c**, Measured responsivities of the 2 × 15 μm² device under TE and TM polarization states, along with the corresponding PDL, obtained at at different vertical positions of the fiber. **d**-**f**, Frequency response from a heterodyne measurement set-up of the 2 × 7 μm² device at 1.5 mA photocurrent (**d**), 2 × 10 μm² device at 2.2 mA photocurrent (**e**) and 2 × 15 μm² device at 3 mA photocurrent (**f**) under -1 V bias. The corresponding 3-dB bandwidths reach 278 GHz, 243 GHz and 206 GHz, respectively. The RF signal is collected by the Rohde & Schwarz power meter (NRP2) in the 0–110 GHz range (hollow circles) and by the VDI power meter (PM5B) in the 110–325 GHz range (solid circles). **g**, RF power versus photocurrent for the MUTC-PDs with different lengths at 180 GHz under -1.5 V bias. **h**, RF power versus photocurrent for the 2 × 15 μm² device at 220 GHz, measured from -1 V to -1.8 V.

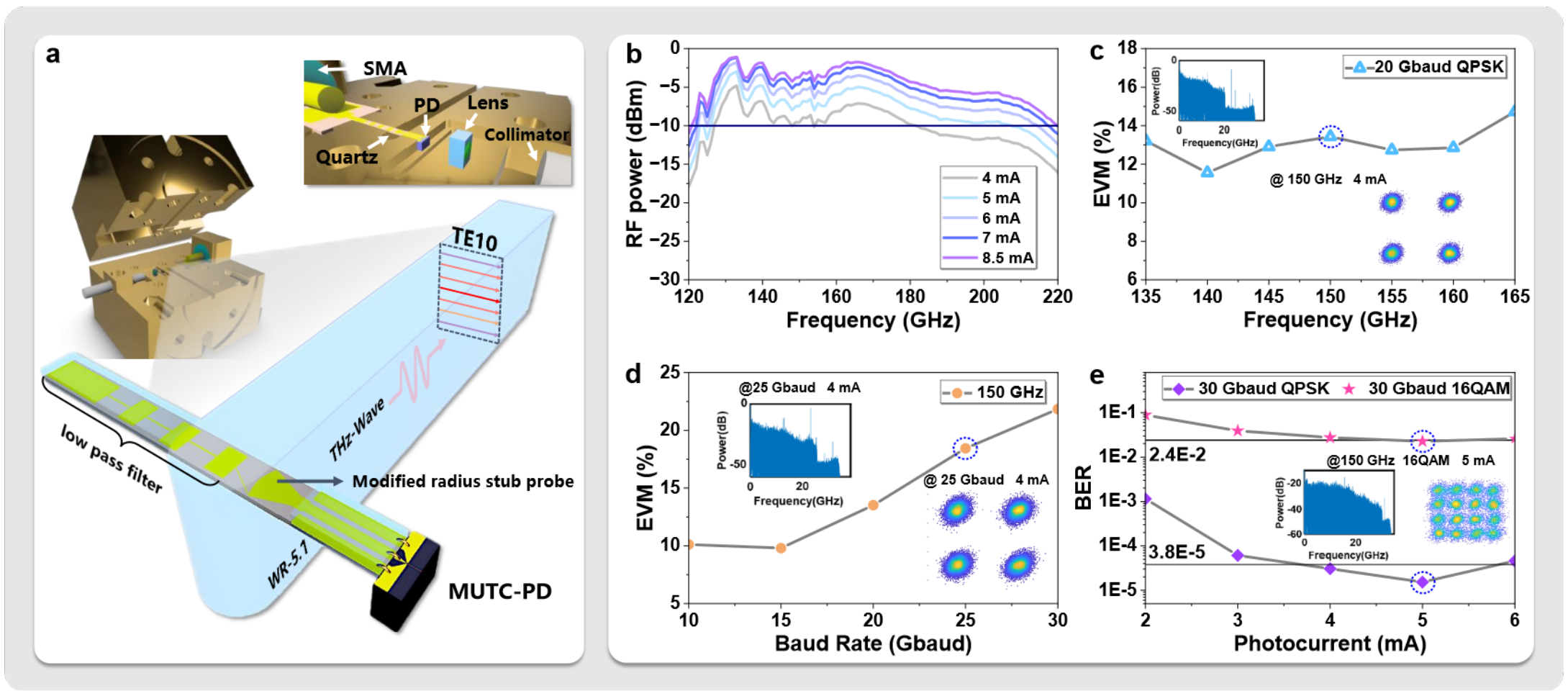

**Fig. 3 WR-5.1 waveguide-packaged MUTC-PD module and the results of photonics-aided sub-THz communication without a THz amplifier.** **a**, Schematic of the WR-5.1 waveguide-packaged MUTCPD module. **b**, Frequency response of the packaged module under different photocurrents from 120 to 220GHz. **c**, EVM versus transmission frequency for a 20Gbaud QPSK signal at a photocurrent of 4mA. Insets: electrical spectrum (left) and constellation diagram (right) at 150GHz. **d**, EVM versus baud rate for a 150GHz carrier frequency at a photocurrent of 4mA. Insets: electrical spectrum (left) and constellation diagram (right) for a 25Gbaud QPSK signal. **e**, BER versus photocurrent for 30Gbaud QPSK/16-QAM signal at a carrier frequency of 150GHz. Insets: electrical spectrum (left) and constellation diagram (right) for a 16-QAM signal at 5mA.

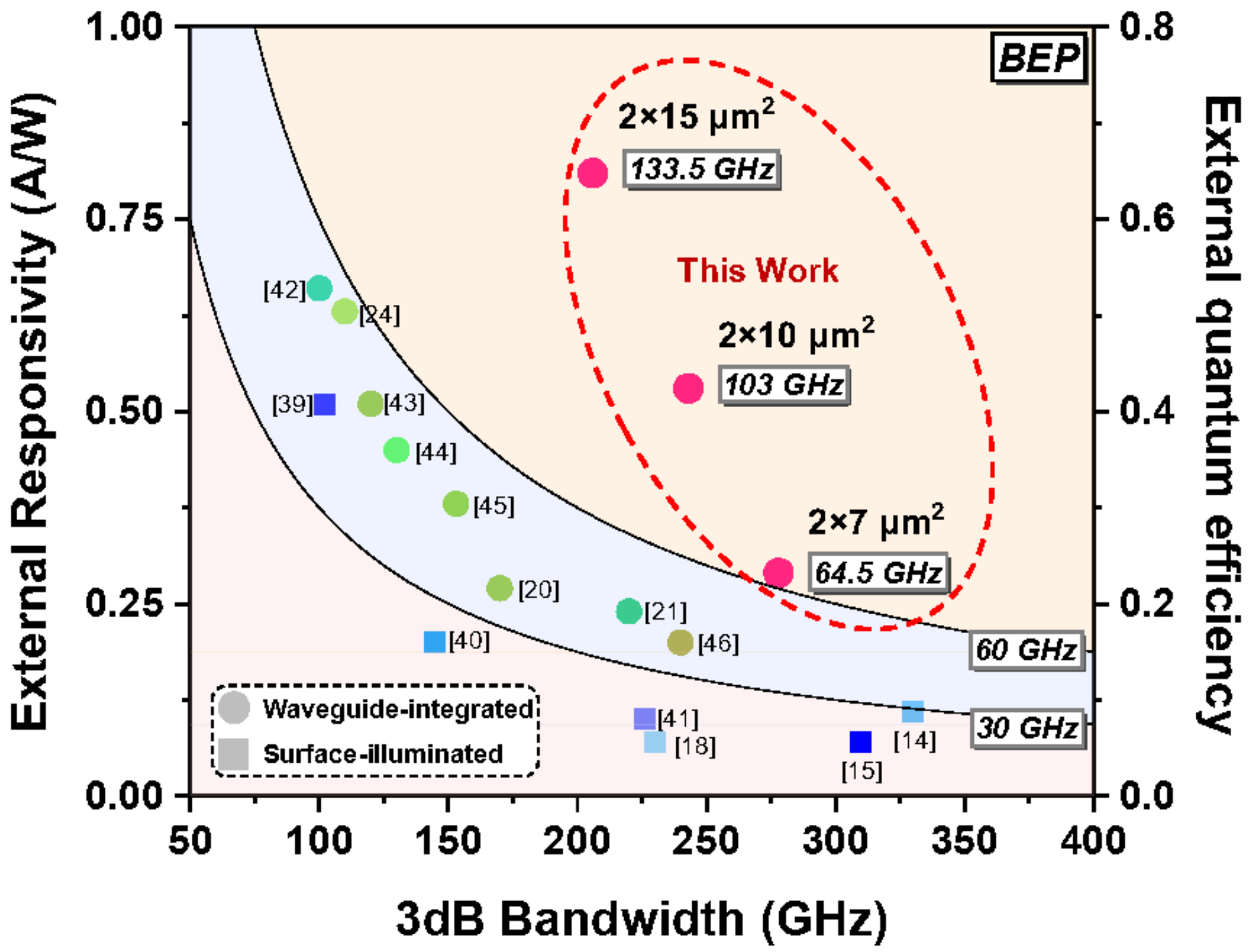

**Fig. 4 The performance comparison of waveguide-coupled and surface illuminated PDs**. 3dB bandwidths and external responsivities of waveguide coupled and surface-illuminated PDs reported in literature[14, 15, 18, 20, 21, 24, 39–46].

# References

[1] Nagatsuma, T., Ducournau, G. & Renaud, C. C. Advances in terahertz communications accelerated by photonics. *Nature Photonics* **10**, 371–379 (2016).

[2] Sung, M. *et al.* Design considerations of photonic THz communications for 6G networks. *IEEE Wireless Communications* **28**, 185–191 (2021).

[3] Zhang, H. *et al.* Tbit/s multi-dimensional multiplexing THz-over-fiber for 6G wireless communication. *Journal of Lightwave Technology* **39**, 5783–5790 (2021).

[4] Kang, G. *et al.* Frequency comb measurements for 6G terahertz nano/microphotonics and metamaterials. *Nanophotonics* **13**, 983–1003 (2024).

[5] Zhang, L., Pang, X., Jia, S., Wang, S. & Yu, X. Beyond 100 Gb/s Optoelectronic Terahertz Communications: Key Technologies and Directions. *IEEE Communications Magazine* **58**, 34–40 (2020).

[6] Seddon, J. P. *et al.* Photodiodes for Terahertz Applications. *IEEE Journal of Selected Topics in Quantum Electronics* **28**, 1–12 (2022).

[7] Ishibashi, T., Muramoto, Y., Yoshimatsu, T. & Ito, H. Unitraveling-Carrier Photodiodes for Terahertz Applications. *IEEE Journal of Selected Topics in Quantum Electronics* **20**, 79–88 (2014).

[8] Ummethala, S. *et al.* THz-to-optical conversion in wireless communications using an ultra-broadband plasmonic modulator. *Nature photonics* **13**, 519–524 (2019).

[9] Zhu, M. *et al.* Ultra-wideband fiber-THz-fiber seamless integration communication system toward 6G: architecture, key techniques, and testbed implementation. *Science China Information Sciences* **66**, 113301 (2023).

[10] Ishibashi, T. *et al.* *Uni-Traveling-Carrier Photodiodes*, UC3 (Optica Publishing Group, 1997).

[11] Ishibashi, T. & Ito, H. Uni-traveling-carrier photodiodes. *Journal of Applied Physics* **127** (2020).

[12] Ishibashi, T. & Ito, H. Uni-Traveling Carrier Photodiodes: Development and Prospects. *IEEE Journal of Selected Topics in Quantum Electronics* **28**, 1–6 (2022).

[13] Ishibashi, T. *et al.* InP/InGaAs uni-traveling-carrier photodiodes. *IEICE transactions on electronics* **83**, 938–949 (2000).

[14] Wun, J.-M., Wang, Y.-W. & Shi, J.-W. Ultrafast Uni-Traveling Carrier Photodiodes With $GaAs_{0.5}Sb_{0.5}/In_{0.53}Ga_{0.47}As$ Type-II Hybrid Absorbers for High-Power Operation at THz Frequencies. *IEEE Journal of Selected Topics in Quantum Electronics* **24**, 1–7 (2018).

[15] Ito, H., Furuta, T., Kodama, S. & Ishibashi, T. InP/InGaAs uni-travelling-carrier photodiode with 310 GHz bandwidth. *Electronics Letters* **36**, 1 (2000).

[16] Huang, Y.-C., Chen, N.-W., Wu, Y.-K., Naseem & Shi, J.-W. Improvements in the Maximum THz Output Power and Responsivity in Near-Ballistic Uni-Traveling-Carrier Photodiodes With an Undercut Collector. *Journal of Lightwave Technology* **42**, 2362–2370 (2024).

[17] Wei, C. *et al.* 150 GHz High-Power Photodiode by Flip-Chip Bonding. *Journal of Lightwave Technology* **41**, 7238–7244 (2023).

[18] Tian, Y. *et al.* Ultrafast MUTC photodiodes over 200 GHz with high saturation power. *Opt. Express* **31**, 23790–23800 (2023).

[19] Li, Q. *et al.* High-Power Evanescently Coupled Waveguide MUTC Photodiode With >105-GHz Bandwidth. *Journal of Lightwave Technology* **35**, 4752–4757 (2017).

[20] Rouvalis, E. *et al.* 170 GHz uni-traveling carrier photodiodes for InP-based photonic integrated circuits. *Opt. Express* **20**, 20090–20095 (2012).

[21] Li, L. *et al.* Ultra-Fast Waveguide MUTC Photodiodes Over 220 GHz. *Journal of Lightwave Technology* **42**, 7451–7457 (2024).

[22] Rouvalis, E., Renaud, C. C., Moodie, D. G., Robertson, M. J. & Seeds, A. J. Continuous Wave Terahertz Generation From Ultra-Fast InP-Based Photodiodes. *IEEE Transactions on Microwave Theory and Techniques* **60**, 509–517 (2012).

[23] Grzeslo, M. *et al.* High saturation photocurrent THz waveguide-type MUTC-photodiodes reaching mW output power within the WR3.4 band. *Optics Express* **31**, 6484–6498 (2023).

[24] Kato, K. *et al.* 110-GHz, 50%-efficiency mushroom-mesa waveguide pin photodiode for a 1.55-$\mu$m wavelength. *IEEE Photonics Technology Letters* **6**, 719–721 (1994).

[25] Runge, P. *et al.* *Broadband 145GHz Photodetector Module Targeting 200GBaud Applications*, M2A.1 (Optica Publishing Group, 2020).

[26] Demiguel, S. *et al.* Very high-responsivity evanescently coupled photodiodes integrating a short planar multimode waveguide for high-speed applications. *IEEE Photonics Technology Letters* **15**, 1761–1763 (2003).

[27] Li, Z., Pan, H., Chen, H., Beling, A. & Campbell, J. C. High-Saturation-Current Modified Uni-Traveling-Carrier Photodiode With Cliff Layer. *IEEE Journal of Quantum Electronics* **46**, 626–632 (2010).

[28] Maloney, T. J. & Frey, J. Transient and steady-state electron transport properties of GaAs and InP. *Journal of Applied Physics* **48**, 781–787 (1977).

[29] Li, Q. *et al.* High-Power Flip-Chip Bonded Photodiode With 110 GHz Bandwidth. *Journal of Lightwave Technology* **34**, 2139–2144 (2016).

[30] Gardes, C., Justice, J., Gity, F., Yang, H. & Corbett, B. *Numerical simulations with energy balance model for unitraveling-carrier photodiode*, 350–353 (2015).

[31] Morgan, J. S. *et al.* Bias-Insensitive GaAsSb/InP CC-MUTC Photodiodes for mmWave Generation up to 325 GHz. *J. Lightwave Technol.* **41**, 7092–7097 (2023).

[32] Chao, E. *et al.* D-Band MUTC Photodiodes With Flat Frequency Response. *IEEE Journal of Selected Topics in Quantum Electronics* **28**, 1–8 (2022).

[33] Guen, A., Sayah, C., Bouazza, B. S. & Sari, N. Steady-State and Transient Electron Transport within Bulk InAs, InP and GaAs: An Updated Semiclassical Three-Valley Monte Carlo Simulation Analysis. *Journal of Modern Physics* **4**, 616-621 (2013).

[34] Latzel, P. *et al.* Generation of mW Level in the 300-GHz Band Using Resonant-Cavity-Enhanced Unitraveling Carrier Photodiodes. *IEEE Transactions on Terahertz Science and Technology* **7**, 800–807 (2017).

[35] Ito, H., Ito, T., Muramoto, Y., Furuta, T. & Ishibashi, T. Rectangular waveguide output unitraveling-carrier photodiode module for high-power photonic millimeter-wave generation in the F-band. *Journal of lightwave technology* **21**, 3456 (2003).

[36] Furuta, T. *et al.* D-band rectangular-waveguide-output uni-travelling-carrier photodiode module. *Electronics Letters* **41**, 1 (2005).

[37] Ito, H., Furuta, T., Muramoto, Y., Ito, T. & Ishibashi, T. Photonic millimetre-and sub-millimetre-wave generation using J-band rectangular-waveguide-output uni-travelling-carrier photodiode module. *Electronics Letters* **42**, 1424–1425 (2006).

[38] Kurokawa, T., Ishibashi, T., Shimizu, M., Kato, K. & Nagatsuma, T. Over 300 GHz bandwidth UTC-PD module with 600 GHz band rectangular-waveguide output. *Electronics Letters* **54**, 705–706 (2018).

[39] Xiong, B. *et al. Ultra-Wide Bandwidth and High Saturation Power Uni-Traveling Carrier Photodiodes*, Tu3D.4 (Optica Publishing Group, 2024).

[40] Beling, A. *et al. High-Speed Integrated Photodiodes*, 1–3 (IEEE, 2019).

[41] Umezawa, T. *et al. DC–226 GHz well-impedance-matched high-speed photoreceiver for multi-band signal detection*, Tu3D.2 (Optica Publishing Group, 2024).

[42] Bach, H.-G. *et al.* InP-based waveguide-integrated photodetector with 100-GHz bandwidth. *IEEE Journal of Selected Topics in Quantum Electronics* **10**, 668–672 (2004).

[43] Beling, A., Bach, H.-G., Mekonnen, G., Kunkel, R. & Schmidt, D. Miniaturized waveguide-integrated p-i-n photodetector with 120-GHz bandwidth and high responsivity. *IEEE Photonics Technology Letters* **17**, 2152–2154 (2005).

[44] Runge, P., Zhou, G., Ganzer, F., Mutschall, S. & Seeger, A. *Waveguide integrated InP-based photodetector for 100Gbaud applications operating at wavelengths of 1310nm and 1550nm*, 1–3 (IEEE, 2015).

[45] Sun, M. *et al.* Ultrafast evanescently coupled waveguide MUTC-PDs with high responsivity. *Opt. Express* **32**, 16455–16466 (2024).

[46] Lischke, S. *et al.* Ultra-fast germanium photodiode with 3-dB bandwidth of 265 GHz. *Nature Photonics* **15**, 925–931 (2021).

[47] Zhou, W. *et al.* 135-GHz D-band 60-Gbps PAM-8 wireless transmission employing a joint DNN equalizer with BP and CMMA. *Journal of Lightwave Technology* **38**, 3592–3601 (2020).

[48] Dat, P. T., Inagaki, K., Kanno, A. & Akahane, K. *60-Gb/s D-Band Wireless Signal Generation and Transmission Using Photonic Technology*, 1–5 (IEEE, 2022).

[49] Wang, K. *et al.* Complex-valued 2D-CNN equalization for OFDM signals in a photonics-aided MMW communication system at the D-band. *Journal of Lightwave Technology* **40**, 2791–2798 (2022).

[50] Zhao, L., Wang, K. & Zhou, W. Transmission of 4096-QAM OFDM at D-band. *Optics Express* **31**, 2270–2281 (2023).

[51] Chen, Q. *et al.* Photonics-aided D-band 64-QAM MMW transmission utilizing modified multi-symbol output neural network equalization. *Optical Fiber Technology* **87**, 103925 (2024).

[52] Tian, Y. *et al.* Photonics-assisted THz wireless communication enabled by wide-bandwidth packaged back-illuminated modified uni-traveling-carrier photodiode. *Opto-Electronic Science* **3**, 230051–1 (2024).

[53] Tian, Y. *et al.* High-Power MUTC Photodiode Module for Photonics-Assisted Beyond 100 Gb/S Wireless Sub-THZ Communications in the F-Band. *Journal of Lightwave Technology* **42**, 5616–5623 (2024).

# 4 Methods

## 4.1 Device fabrication

The epitaxial structure of the designed MUTC-PD was grown by a metal-organic chemical vapor deposition (MOCVD) system on a 2-inch semi-insulating InP substrate. The device was fabricated into a triple-mesa structure. Inductively coupled plasma (ICP) dry etching process was conducted to achieve vertical sidewalls and control of mesa dimensions, while wet etching ($H_3PO_4$:HCl = 3:1) was used for electrical isolation between n-mesas. Ti/Pt/Au and GeAu/Ni/Au layers were deposited to form the p-contact and n-contact, respectively, with a rapid thermal annealing process conducted at 360 °C . After passivating the sidewalls with a BCB layer, Ti/Au layers were deposited to create CPW pads. Detailed fabrication process flow can be found in the Supplementary Note 1.

## 4.2 DC electrical measurements

The dark current was measured with Keysight B1500 semiconductor parameter analyzer at room temperature. The responsivity of the device was characterized with a laser source (Keysight 81606A) and a source measurement unit (SMU, Keysight B2901A). Light from a cleaved single-mode fiber was coupled to the waveguide through an automated fiber alignment process. Polarization was manually optimized using a polarization controller. The laser wavelength was set to 1550 nm, and the power at the fiber tip was measured using an optical power meter (Thorlabs PM400). External responsivity was calculated as the ratio of the photocurrent to the optical power at the fiber tip, under the desired voltage. To characterize the PDL of the device, a Keysight N7786B polarization synthesizer was employed to change and stabilize the polarization state of the incident light.

## 4.3 RF measurements

The frequency response and the saturation performance of the InGaAs/InP waveguide devices and packaged module were investigated by an optical heterodyne setup. Two tunable lasers with wavelengths near 1.55 μm were combined to generate a heterodyne beat note with a wide frequency tuning range from Megahertz to Terahertz.

For PD chips, the output RF power was obtained by two measurement setups. From DC to 110 GHz, the RF signal was collected by the Rohde & Schwarz power meter (NRP2) through the GSG Model 110H probe. As frequencies exceed 110 GHz, a VDI power meter (PM5B) and a set of waveguide probes covering 110 GHz to 325 GHz were used to measure the THz power generation. Similarly, since the packaged module featured a standard waveguide interface, the output RF power can be easily obtained by a matched waveguide taper and VDI power meter (PM5B). The detailed schematics of two measurement setups can be founded in Supplementary Note 4. After extracting the S-parameters of the probe and bias tee, and measuring the reflection coefficient of the PD, the actual RF power delivered by the PDs was calculated using the equations provided in Supplementary Note 5.

## Data availability

All data supporting the findings of this study are provided within the Article. The Source Data are available via figshare at https://doi.org/10.6084/m9.figshare.28659041(ref. [54]). Additional data related to this work are available from the corresponding authors upon reasonable request.

## References

[54] Li, L. Supporting Data for article Over-200 GHz bandwidth MUTC Photodiodes with0.81 A/W external responsivity. figshare, (2025).